\documentstyle[12pt]{article}
\newcommand{\beq}{\begin{equation}}
\newcommand{\eeq}{\end{equation}}

\newcommand{\sg}{\sigma}

\renewcommand{\b}{\beta}
\renewcommand{\a}{\alpha}

\newcommand{\th}{\theta}

\newcommand{\r}{\prime}

\begin{document}
\topmargin 0pt
\oddsidemargin 5mm
\headheight 0pt
\headsep 0pt
\topskip 5mm

\begin{flushright}
BCUNY-HEP-96-1\\
\hfill
January 1996
\end{flushright}

\begin{center}
\hspace{10cm}

\vspace{48pt}
{\large \bf
LATTICE BOSONIZATION}
\end{center}

\vspace{15pt}

\begin{center}
{\bf Maxime Kudinov}$^{1,2}$

\vspace{10pt}
and
\vspace{10pt}

{\bf Peter Orland}$^{2,3}$

\end{center}

\vspace{40pt}

\begin{center}
{\bf Abstract}

\end{center}

A free lattice fermion field theory
in 1+1 dimensions can be interpreted as SOS-type model, whose
spins are integer-valued. We point out that the relation
between these spins and the fermion field is similar to the abelian
bosonization relation between bosons and fermions
in the continuum. Though on the lattice
the connected $2n$-point correlation
functions of the integer-valued
spins are not zero for any $n \ge 1$, the two-point correlation
function of these spins is that of free bosons in the infrared. We
also conjecture the form of the Wess-Zumino-Witten chiral field
operator in a nonabelian lattice fermion model. These constructions
are similar in spirit to the ``twistable string"
idea of Krammer and Nielsen.

\vspace{40pt}

\noindent
----------------------------------------------------------------------------

\noindent
$^{1}$ Physics Department, Queens
College, The City University of New York, Flushing, NY 11367

\noindent
$^{2}$The Graduate School and University Center, The
City University of New York, 33 W. 42nd St., New York, NY 10036

\noindent
$^{3}$Physics Department, Baruch College, The
City University of New York, New York, NY 10010

\vfill
\newpage

\section{Introduction}

The observation that
boson excitations exist in theories of fermions in one
space and one time dimension was first made by Tomonaga
\cite{tom}. Subsequent advances were made for the Luttinger
model \cite{math} and the massive Thirring
model \cite{sid}.

The relation of the gradient of
a one-component boson field $\phi$ and the current
of a one-flavor
dirac fermion field $\psi$ is
\beq
:{\bar \psi} \gamma_{\mu} \psi :
= \pi^{-\frac{1}{2}} \epsilon_{\mu \nu} \partial^{\nu} \phi\;, \label{1}
\eeq
where $\mu, \nu=0$ is the time index and $\mu, \nu=1$ is the space index.
One can show that the local commutation
relations of the operators on each side of equation
(\ref{1}) are the same. This relation is not easily
generalized to the lattice where the algebra of the local
fermion
currents does not close. In fact, one does not expect a precise
lattice analogue of (\ref{1}). The best one can hope for is that a ``good"
set of mutually
commuting
operators can be defined in a free
lattice fermion theory. Such a set of operators would be
good in the sense that their correlation functions become those of
free boson
fields in the infrared. In other words, the boson correlation
functions
should flow along renormalization-group
trajectories to the correlation functions of the
gaussian model.

Integrating equation (\ref{1}) for $\mu =0$ gives
\beq
\phi(x)=:{\sqrt \pi} \int_{0}^{x} dz [\psi^{\dag}(z) \psi(z)]: \;,
\label{1.1}
\eeq
where one boundary is at $x=0$. The
constant of integration is fixed
to zero by the
requirement that the vaccumm
expectation value
of $\phi$ is zero. We will show that there is
a natural lattice analogue of (\ref{1.1}).

A free lattice fermion field theory
in 1+1 dimensions can be interpreted as SOS-type model, whose
spins are integer-valued. We point out that the relation
between these spins and the fermion field is similar to the
bosonization relation between bosons and fermions
in the continuum. While the connected $2n$-point correlation
functions of the integer-valued
spins are not zero for any $n \ge 1$, the two-point correlation
function of these spins is that of free bosons in the infrared. This
supports the arguments of den Nijs that in the massless phase, SOS-type
models are described in the infrared
by the gaussian model \cite{dn}.

We begin by considering the antiferromagnetic
XX chain \cite{liebetal}, with hamiltonian
\beq
H=\frac{1}{2}
\sum_{n=1}^{N-1} \sum_{\pm} \sg^{\pm}_{n} \sg^{\mp}_{n+1}\;,  \label{2}
\eeq
where the spin operators at each lattice site, $n$, are
pauli matrices $\sg^{x}_{n}$, $\sg^{y}_{n}$ and
$\sg^{z}_{n}$, with $2\sg^{\pm}_{n}=\sg^{x}_{n} \pm i\sg^{y}_{n}$.

The XX chain can be viewed
as a lattice model of free relativistic
fermions. This can
be seen by making a Jordan-Wigner
transformation \cite{frad} to Fermion fields
\beq
\psi_l= \left(  \begin{array}{c}  \a_{l}  \\
                                   \b_{l}
        \end{array} \right) \;,\;\;
\psi^{\dag}_l= \left(  \begin{array}{c}  \a^{\dag}_{l}  \\
                                          \b^{\dag}_{l}
        \end{array} \right) \;,   \label{2.1}
\eeq
where
\begin{eqnarray}
\a^{\dag}_{l}=\sg^{+}_{2l+1} \prod_{m=1}^{2l} (-i\sg^{z}_{m})\;,\;\;
\a_{l}=\sg^{-}_{2l+1} \prod_{m=1}^{2l} (i\sg^{z}_{m})\;, \nonumber \\
\b^{\dag}_{l}=\sg^{+}_{2l} \prod_{m=1}^{2l-1} (-i\sg^{z}_{m})\;,\;\;
\b_{l}=\sg^{-}_{2l} \prod_{m=1}^{2l-1} (i\sg^{z}_{m})\;, \label{2.2}
\end{eqnarray}
which satisfy  the local anticommutation relations
\beq
[\a^{\dag}_{l}, \a_{l^{\r}}]_{+}=\delta_{l l^{\r}}\;,\;\;
[\b^{\dag}_{l}, \b_{l^{\r}}]_{+}=\delta_{l l^{\r}}\;,  \label{2.3}
\eeq
with all other anticommutators equal to zero. It is easy to transform
(\ref{2}) into a dirac hamiltonian with these operators. A naive lattice
operator resembling (\ref{1.1}) is
\beq
\phi_{n}={\sqrt \pi} \sum_{l=0}^{n}  : \psi^{\dag}_{l} \psi_{l} :
={\sqrt \pi} \sum_{l=0}^{n}  (\frac{1}{2}\sg^{z}_{2l}
+\frac{1}{2}\sg^{z}_{2l+1})\;. \label{2.4}
\eeq
It is well known that the vaccum
expectation value of $\sg^{z}_{n}$ vanishes \cite{liebetal}. We
will show that the natural
choice of a lattice
bosonic field operator is similar, though
not identical to (\ref{2.4}). Both operators become
(\ref{1.1}) in the continuum limit.

It is also true that a spin configuration can be viewed
as a ``height profile" through the integer-eigenvalued
operator
\beq
h_{n} = \sum_{l=1}^{n} \sg^{z}_{l}\;.      \label{3}
\eeq
This operator strongly resembles $\phi_{l}$ defined
in (\ref{2.4}) (except
for a factor of $\frac{{\sqrt \pi}}{2}$). The
eigenvalues ${\lambda}_{n}$ of the
$h_{n}$ are restricted by
${\lambda}_{n+1}-{\lambda}_{n}=0, \pm 1$, ${\lambda}_{0}=0$. The
Hamiltonian (\ref{1}) is equivalent to a restricted
SOS-model hamiltonian
\beq
H=\frac{1}{2}
\sum_{n=1}^{N-1} \sum_{\pm} h^{\pm}_{n} h^{\mp}_{n+1}\;,  \label{4}
\eeq
where
\beq
[h^{\pm}_{n}, h_{m}]= \mp  h^{\pm}_{n} \delta_{n m}\;, \;\;
[h^{+}_{n},h^{-}_{m}]=0 \;.              \label{5}
\eeq
In the ground state of the XX chain, the
magnetization
is zero. Therefore it is clear that the expectation value
of $h_{n}$ is zero. One can imagine ``coarse graining" the
spin configurations of the SOS model by some sort of renormalization
procedure. Since the model is known to be gapless, it is reasonable
to conjecture that in
the infrared the height
profile
becomes a continuous
function, described by
a free massless boson field theory  of
a field $\phi$ which vanishes on the boundaries. One
piece
of supporting evidence for this
assertion is that the central
charge of the conformally invariant
critical
theory can be found by
calulating the casimir energy (the connection between
the two can be found in \cite{affcar}). The
casimir energy turns
out to be identical to that for free bosons with
dirichlet boundary conditions.

From the above discussion, it seems that
the lattice boson field should be
\beq
\phi_{n} = \frac{{\sqrt \pi}}{2} h_{n}=
\frac{{\sqrt \pi}}{2} \sum_{l=1}^{n} \sg^{z}_{l}\;.     \label{5.1}
\eeq
In
this letter we will show that this is indeed the case.

By a different jordan-wigner transformation \cite{liebetal}
\beq
c^{\dag}_{n}=\sg^{+}_{n} \prod_{m=1}^{n-1} \sg^{z}_{m}\;,\;\;
c_{n}=\sg^{-}_{n} \prod_{m=1}^{n-1} \sg^{z}_{m}\;, \label{6}
\eeq
which implies the local anticommutation relations
\beq
[c^{\dag}_{n}, c_{m}]_{+}=\delta_{n m}\;,\;\;
[c_{n}, c_{m}]_{+}=[c^{\dag}_{n}, c^{\dag}_{m}]_{+}=0 \;,  \label{7}
\eeq
the antiferromagnetic H=hamiltonian becomes a hopping hamiltonian
of free fermions
\beq
H=\sum_{k} \cos k \;c^{\dag}(k)c(k)    \;, \label{8}
\eeq
where $k=\frac{\pi}{N+1} m$, $\;m=1,...,N$ and
\beq
c_{n}=(\frac{2}{N})^{\frac{1}{2}} \sum_{k} \sin nk \; c(k)\;,\;\;
c^{\dag}_{n}
=(\frac{2}{N})^{\frac{1}{2}} \sum_{k} \sin nk \; c^{\dag}(k)\;. \label{9}
\eeq
In this formulation, the two-point correlation function
of $\phi_{n}$ in the heisenberg representation is
\beq
f(n,t;n^{\r}, t^{\r})=<0| {\cal T}
\{ \phi^{H}_{n}(t) \phi^{H}_{n^{\r}}(t^{\r}) \} |0>
=\pi
<0| {\cal T} \{ h^{H}_{n}(t) h^{H}_{n^{\r}}(t^{\r}) \} |0> \;, \label{10}
\eeq
where for any operator
$A$, $A^{H}(t) \equiv e^{-iHt} A e^{iHt}$. In terms of
the operators
(\ref{8}) this two-point function is
\beq
f(n,t;n^{\r}, t^{\r})
=\frac{\pi}{4}
<0| {\cal T} \{ \sum_{l=1}^{n} [2c^{\dag\;H}_{l}(t)\,c^{H}_{l}(t)-1]
\sum_{l^{\r}=1}^{n^{\r}}
[2c^{\dag\;H}_{l^{\r}}(t^{\r})\,c^{H}_{l^{\r}}(t^{\r})-1] \} |0>
\;. \label{11}
\eeq

We will calculate $f(n,t;n^{\r}, t^{\r})$ for $t=t^{\r}$
exactly and show that
for $N>>|n-n^{\r}|>>1$ it has the form of
the two-point function of a free massless
bosonic
field theory, in
which the boson field vanishes at the boundaries. For
$t\neq t^{\r}$, the lorentz transformation
properties
of the low-lying
states
should imply that the two-point function
has the standard form in
this case as well. We will give some further arguments
supporting this assertion
at the end of this letter.

We feel it is important to point out that while (\ref{10}) is
the correlation function of a bosonic theory which
becomes free in the infrared, {\em the
theory is not free on the lattice}. Indeed, the higher-point
lattice green's functions $<0| {\cal T}
\{ \phi^{H}_{n_{1}}(t)\;.\;.\;.
\phi^{H}_{n_{2M}}(t^{\r}) \} |0>$, $M>1$, are
not zero. It seems clear that the
interaction
in the bosonic theory is irrelevant and that all these
higher-point functions vanish in the infrared. However, we
have not yet systematically studied this issue.

The expression (\ref{11}) can be reduced to
\beq
f(n,t;n^{\r}, t^{\r})
=-\frac{4\pi}{N^{2}} \sum_{l=1}^{n} \sum_{l^{\r}=1}^{n^{\r}}
\sum_{k, k^{\r}} G(k, t^{\r}-t)G(k^{\r}, t-t^{\r})
\sin kl \sin kl^{\r} \sin k^{\r}l \sin k^{\r}l^{\r}\;,  \label{12}
\eeq
where the minus sign results from fermi statistics and the standard
fermion
propagator, $G(k,t-t^{\r})$, is given by
\begin{eqnarray}
G(k,t-t^{\r})&=&<0| {\cal T}
              \{ c^{H}(k,t) c^{\dag\;H} (k^{\r},t^{\r}) \} |0> \nonumber \\
             &=&[\th (t-t^{\r}) <0|c(k)c^{\dag}(k)|0>
                 -\th (t^{\r}-t)
<0|c^{\dag}(k)c(k)|0>] e^{-i(t-t^{\r})\cos k }\;.   \nonumber \\
\label{13}
\end{eqnarray}
Note that $<0|c^{\dag}(k)c(k)|0>=\th(-\cos k)$ and $<0|c(k)c^{\dag}(k)|0>
=\th(\cos k)$ (this follows from the fact that
the fermi surface is at $\cos k=0$).

Using
\beq
\sum_{l=1}^{n} \sin kl \sin k^{\r} l \;
\sum_{l^{\r}=1}^{n^{\r}} \sin kl^{\r} \sin k^{\r}l^{\r}
=\frac{1}{8} \sum_{s=-n}^{n}\sum_{s^{\r}=-n^{\r}}^{n^{\r}}
e^{ik(s+s^{\r})}
[\cos k^{\r}(s+s^{\r})-  \cos k^{\r}(s-s^{\r})] \;. \label{14}
\eeq
Neglecting a contribution of $O(1/N^{2})$ from the fermi surface gives
\begin{eqnarray}
f(n,t;n^{\r}, t^{\r})
&=&-\frac{\pi}{2N^{2}} \sum_{s=-n}^{n} \sum_{s^{\r}=-n^{\r}}^{n^{\r}}
         [\; \th(t^{\r}-t) \sum_{m=1}^{\frac{N}{2}}\;-
        \; \th(t-t^{\r})\sum_{m=\frac{N}{2}+1}^{N}] \nonumber \\
&\times&  [\; \th(t-t^{\r}) \sum_{m^{\r}=1}^{\frac{N}{2}}\;-\;
          \th(t^{\r}-t)\sum_{m^{\r}=\frac{N}{2}+1}^{N}]
         e^{i\frac{\pi}{N+1}(s+s^{\r})m}        \nonumber \\
&\times& [\cos \frac{\pi}{N+1}(s+s^{\r})m-
          \cos \frac{\pi}{N+1}(s-s^{\r})m^{\r}]  \nonumber \\
&\times&
\exp [ i(t-t^{\r}) (\cos \frac{\pi m}{N+1} - \cos \frac{\pi m^{\r}}{N+1}) ]\;.
\label{15}
\end{eqnarray}
At equal times, $t=t^{\r}$, this expression simplifies further, so that
we may carry out the summation over $m$ and $m^{\r}$. Replacing
$N+1$ by $N$ introduces errors of order $1/N$ and will not affect
(\ref{15}) in the thermodynamic limit. The result is
\begin{eqnarray}
f(n, n^{\r})
&\equiv& f(n,t;n^{\r}, t)=\frac{i\pi}{2N^{2}}
         \sum_{s=-n}^{n} \sum_{s^{\r}=-n^{\r}}^{n^{\r}}
         \frac{\sin^{2} \frac{\pi}{4} (s+s^{\r}) e^{i\frac{\pi}{2}(s+s^{\r})}}
         {\sin \frac{\pi}{2N} (s+s^{\r})} \nonumber \\
&\times& [\;
\frac{\sin^{2}\frac{\pi}{4} (s+s^{\r}) \sin \frac{\pi}{2}(s+s^{\r})}
{\sin\frac{\pi}{2N} (s+s^{\r})}
-\frac{\sin^{2}\frac{\pi}{4} (s-s^{\r}) \sin \frac{\pi}{2}(s-s^{\r})}
{\sin\frac{\pi}{2N} (s-s^{\r})}\;]\;. \label{16}
\end{eqnarray}
The non-vanishing terms in the sum on the right-hand-side of
(\ref{16}) are those for which $s+ s^{\r}$ is odd. For large
$N$ we may replace this sum by an integral, dividing by a factor of
two (because half of the terms in the sum vanish). Defining
$x=\frac{\pi s}{2N}$ and
$x^{\r}=\frac{\pi s^{\r}}{2N}$, the correlation function
is
\beq
f(n, n^{\r})=\frac{1}{8\pi}
\int_{-\frac{\pi n}{2N}}^{\frac{\pi n}{2N}} dx
\int_{-\frac{\pi n^{\r}}{2N}}^{\frac{\pi n^{\r}}{2N}} dx^{\r} \;
[\; \frac{1}{\sin^{2} (x+x^{\r})}-\frac{1}{\sin (x+x^{\r})
\sin (x-x^{\r})} \;]\;. \label{17}
\eeq
Evaluation of the integral yields
\beq
f(n, n^{\r})=\frac{1}{2\pi} [\; \ln |\frac{\sin \frac{\pi}{2N}(n+n^{\r})}
{\sin \frac{\pi}{2N}(n-n^{\r})}|
- Li_{2} (\frac{\tan \frac{\pi n}{2N}}{\tan \frac{\pi n^{\r}}{2N}})
+ Li_{2} (-\frac{ \tan \frac{\pi n}{2N}}
{\tan \frac{\pi n^{\r}}{2N}})\;]\;, \label{18}
\eeq
where the polylogarithm function
$Li_{2}(z)$ is given by
$Li_{2}(z) = \sum_{r=1}^{\infty} \frac{z^{r}}{r^{2}}$
\cite{prudnikov}.

It is simple to take the continuum limit
of (\ref{18}). If we set
$r=na$ and $r^{\r}=n^{\r}a$, where $a$ is the lattice
spacing and take $N\rightarrow \infty$, with volume
$L=Na$ fixed, the
two-point function of bosonic fields is
\beq
f(r, r^{\r})=\frac{1}{2\pi} [\; \ln |\frac{\sin \frac{\pi}{2L}(r+r^{\r})}
{\sin \frac{\pi}{2L}(r-r^{\r})}|
- Li_{2} (\frac{\tan \frac{\pi r}{2L}}{\tan \frac{\pi r^{\r}}{2L}})
+ Li_{2} (-\frac{ \tan \frac{\pi r}{2L}}
{\tan \frac{\pi r^{\r}}{2L}})\;]\;, \label{19}
\eeq
which vanishes for $r=0$ or $r^{\r}=L$.

For $L>>|r-r^{\r}|$, the arguments of the polylogarithm functions
tend to plus one and minus one. Therefore the last two terms merely
give a constant contribution:
\beq
Li_{2} (\frac{\tan \frac{\pi r}{2L}}{\tan \frac{\pi r^{\r}}{2L}})
+ Li_{2} (-\frac{ \tan \frac{\pi r}{2L}}
{\tan \frac{\pi r^{\r}}{2L}})\; \;\; \longrightarrow \;\;
\sum_{r=1}^{\infty} \frac{1-(-1)^{r}}{r^{2}}=\frac{\pi^{2}}{4}\;, \label{19.1}
\eeq
so
\beq
f(r, r^{\r}) \approx
- \frac{1}{2\pi}\ln |\sin \frac{\pi}{2L}(r-r^{\r})| +
\frac{1}{2\pi}\ln |\sin \frac{\pi}{2L}(r+r^{\r})|-\frac{\pi}{8}
\;, \label{19.2}
\eeq
which is the massless
scalar propagator for dirichlet
boundary conditions
with an expected non-universal
constant added. Such constants always appear with
lattice regularizations (see for example Spitzer \cite{spitzer}).

We have only calculated the
equal-time two-point function of
$\phi_{n}$. There is a simple argument
(which carries more weight than
that based on lorentz properties
of the spectrum) that the time-ordered
correlation function
of two field operators at the same lattice site but at different
times should also be that of free bosons in the
infrared. The ground state of the XX chain coincides with
that of the six-vertex model with parameter $\Delta$ equal
to zero \cite{6vertex}. The six-vertex model has an SOS
interpretation similar to that of the spin chain \cite{dn}. Since
the boltzmann factor of this model
has $90^{o}$ rotation invariance, its correlation
functions should have this invariance also. In the continuum
limit, the six-vertex model coincides with the
XX chain, so the two-point
correlation functions
of the latter will also be $90^{o}$ rotation invariant. We
suspect that the two-point function can be calculated in the
case of arbitrary space and time seperation.

We have attempted to make a similar construction on the
lattice for nonabelian
bosonization \cite{witten}. In particular, we
have found a candidate
for the chiral boson field
operator of the Wess-Zumino-Witten
model, for global symmetry group
$spin(4) \bigotimes U(1)$. We have
not evaluated any of
the correlation functions of this field as yet.

If $\psi^{A}$, $A=1,...,4$ is a $1+1$-dimensional
dirac field, the
Wess-Zumino-Witten field, $g \; \varepsilon \; spin(4) \bigotimes U(1)$, is
given by
\beq
{\bar \psi}^{A} \gamma_{\mu} \psi^{B}
= i\frac{1}{2\pi} \epsilon_{\mu \nu} [g \partial^{\nu}
g^{-1}]^{AB} \;, \label{20}
\eeq
Integrating equation (\ref{20}) for $\mu =0$ gives the equation
analogous to (\ref{1.1})
\beq
g(x)={\cal P} \exp 2\pi i\int_{0}^{x} dz \psi^{\dag}(z) \psi(z) \;,
\label{21}
\eeq
where ${\cal P}$ denotes path ordering, one boundary is again
at $x=0$
and the bilinear $\psi^{\dag}(z) \psi(z)$
is understood as a $4 \times 4$ matrix.

There is a lattice spin chain which is equivalent to free
fermions with the symmetry group
$spin(4) \bigotimes U(1)$ \cite{acs}, \cite{or}. Its hamiltonian is
\beq
H=\frac{1}{4}
\sum_{n=1}^{N-1} \{
[1+(-1)^{n}]\gamma^{A}_{n} \gamma^{A}_{n+1}
+[1-(-1)^{n}]\rho^{A}_{n} \rho^{A}_{n+1} \}\;,  \label{22}
\eeq
where the operators at each site $n$ are defined by
\beq
[\gamma^{A}_{n}, \gamma^{B}_{n}]_{+}=2\delta^{AB}\;,\;\;
\gamma^{5}_{n}= \gamma^{1}_{n} \gamma^{2}_{n} \gamma^{3}_{n}
\gamma^{4}_{n}\;,\;\;
\rho^{A}_{n}=-i\gamma^{5}_{n}\gamma^{A}_{n}\;,\;\;
\sg^{AB}_{n}=\frac{-i}{4} [\gamma^{A}_{n},\gamma^{B}_{n}]\;,  \label{23}
\eeq
which constitute a basis for the lie algebra of SU(4). The operators
$\gamma^{A}_{n}$ should
not be confused with the matrices $\gamma_{\mu}$. The $6$ generators
of $spin(4)$ are $\sum_{n}\sg^{AB}_{n}$ while the generator of $U(1)$
is $\sum_{n}\gamma^{5}_{n}$.

A jordan-wigner
transformation \cite{acs}, \cite{or} analogous to (\ref{2.1}), (\ref{2.2})
is
\beq
\psi^{A}_l= \left(  \begin{array}{c}  \a^{A}_{l}  \\
                                   \b^{A}_{l}
        \end{array} \right) \;, \label{24}
\eeq
where
\begin{eqnarray}
\a^{A}_{l}=\frac{1}{{\sqrt 2}}
\gamma^{A}_{2l+1} \prod_{m=1}^{2l} \gamma^{5}_{m}\;,\;\;
\b^{A\;\dag}_{l}=\frac{1}{{\sqrt 2}}
\rho^{A}_{2l} \prod_{m=1}^{2l-1} \gamma^{5}_{m}\;,
\label{25}
\end{eqnarray}
which satisfy  the local anticommutation relations
\beq
[\a^{A}_{l}, \a^{B}_{l^{\r}}]_{+}=\delta^{AB} \delta_{l l^{\r}}\;,\;\;
[\b^{A}_{l}, \b^{B}_{l^{\r}}]_{+}=\delta^{AB} \delta_{l l^{\r}}\;,
\label{26}
\eeq
with all other anticommutators equal to zero. This transformation
converts (\ref{22}) into a free massless
hamiltonian of
a majorana-dirac field with $4$ real components.

The lattice operator-valued matrix
analogous to the continous operator-valued matrix
$\psi_{A}^{\dag}(z) \psi_{B}(z)$ is
\beq
\Omega^{AB}_{n}=\gamma^{A}_{n}\gamma^{B}_{n}=\rho^{A}_{n}\rho^{B}_{n}=
\delta^{AB}+2i\sg^{AB}_{n}\;.   \label{27}
\eeq
On the basis of (\ref{21})
we conjecture that the correlation functions
of
\beq
g_{n}=\prod_{m=1}^{n} \exp 2\pi i \Omega_{m} \;,\label{28}
\eeq
are those of the Wess-Zumino-Witten model in the infrared.

Some time ago Krammer and Nielsen proposed the idea that a
``twistable string" \cite {hbn}, with only bosonic
degrees of freedom on the world-sheet was equivalent to
a Neveu-Schwartz-Ramond string. These authors were
motivated
by the construction in reference
\cite{acs} and argued (using methods
quite different from ours) that one could describe bosons
as well as fermions using $(1+1)$-dimensional spin densities.

In summary, we have found the operator in a $1+1$-dimensional
theory of free
relativistic fermions whose two-point function
is that of free bosons in the infrared. This operator is proportional
to the
integer-valued spin
in an SOS formulation of the fermionic hamiltonian. Finally
we have conjectured
a similar bosonization relation for a nonabelian theory of
relativistic fermions.

\vfill

\end{document}